\documentclass[%
reprint,
showpacs,preprintnumbers,
amsmath,amssymb,
aps,
]{revtex4-1}

\usepackage{subfigure}
\usepackage{wrapfig}
\usepackage{color}
\usepackage{graphicx}
\usepackage{dcolumn}
\usepackage{bm}
\usepackage{txfonts}

\begin{document}


\title{Stability of multinode Dirac semimetals against strong long-range correlations}

\author{Akihiko Sekine}
\email{sekine@imr.tohoku.ac.jp}
\author{Kentaro Nomura}
\affiliation{Institute for Materials Research, Tohoku University, Sendai 980-8577, Japan}

\date{\today}

\begin{abstract}
We study the stability of Dirac semimetals with $N$ nodes in three spatial dimensions against strong $1/r$ long-range Coulomb interactions.
We particularly study the cases of $N=4$ and $N=16$, where the $N=4$ Dirac semimetal is described by the staggered fermions and the $N=16$ Dirac semimetal is described by the doubled lattice fermions.
We take into account the $1/r$ long-range Coulomb interactions between the bulk electrons.
Based on the U(1) lattice gauge theory, we analyze the system from the strong coupling limit.
It is shown that the Dirac semimetals survive in the strong coupling limit when the out-of-plane Fermi velocity anisotropy of the Dirac cones is weak, whereas they change to Mott insulators when the anisotropy is strong.
A Possible global phase diagram of correlated multinode Dirac semimetals is presented.
Implications of our result to the stability of Weyl semimetals and three-dimensional topological insulators are discussed.
\end{abstract}

\pacs{
71.30.+h, 
11.15.Ha, 
03.65.Vf, 
71.35.-y 
}

\maketitle

\section{Introduction}
Dirac semimetals in three spatial dimensions have gapless three-dimensional (3D) linear dispersions, i.e. 3D Dirac cones, in the bulk.
They can be regarded as a 3D analog of graphene.
After the theoretical predictions had been made \cite{Young2012,Wang2012,Wang2013}, the Dirac semimetals such as Na$_3$Bi and Cd$_3$As$_2$ were experimentally discovered recently \cite{Liu2014,Neupane2014,Borisenko2014,Cheng2014,He2014}.
These Dirac semimetals possess two Dirac nodes which are protected by crystalline symmetry.
One of the important meanings of the realization of Dirac semimetals is on the point that they can lead to various topological phases, since they lie next to various topological phases in the phase diagrams.
In 3D topological insulators, the bulk energy gap closing is required to make the system turn into normal band insulators \cite{Fu2007,Fu2007a,Zhang2009}.
At these transition points, Dirac semimetals can be realized.
Experimentally, such a continuous transition is observed in the solid-solution system TlBi(S$_{1-x}$Se$_x$)$_2$ \cite{Sato2011}.
Further, Weyl semimetals can be realized when time-reversal or inversion symmetry breaking occurs in Dirac semimetals \cite{Volovik2003,Volovik2007,Murakami2007,Burkov2011,Kurebayashi2014}.
Regardless of intensive searches, Weyl semimetal phases have not been experimentally observed so far.
Hence, it is expected that recent experimental realization of the Dirac semimetals also gives rise to the realization of Weyl semimetal phases.

Stability of topological phases against electron correlation is one of attractive themes.
It has been shown that strong short-range interactions break 2D topological insulator phases \cite{Rachel2010,Varney2010,Hohenadler2011,Yu2011,Yamaji2011,Zheng2011,Hohenadler2012,Yoshida2012,Tada2012,Hohenadler2013,Liu2013}, 3D topological insulator phases \cite{Pesin2010,Sekine2013}, and Weyl semimetal phases \cite{Wei2012,Sekine2013,Maciejko2013}.
On the other hand, recent studies have suggested that these topological phases can survive strong $1/r$ long-range Coulomb interactions \cite{Sekine2013a,Araki2013,Sekine2013b}.
What about in Dirac semimetals?
In Dirac semimetals, the effects of long-range interactions are expected to be important, since the screening effect is considered to be weak due to the vanishing density of states near the Fermi level. 

Effects of long-range Coulomb interactions in graphene have been studied widely \cite{Kotov2012}.
Monolayer graphene on a substrate with sufficiently small dielectric constant has been predicted theoretically to be insulating (i.e. Dirac fermions become massive) due to strong $1/r$ Coulomb interactions \cite{Khveshchenko2001,Khveshchenko2004,Khveshchenko2007,Hands2008,Drut2009,Drut2009a,Araki2010,Armour2011,Giedt2011,Buividovich2012}.
However, as the number of layers is increased, it has been found that the semimetal phase survives strong $1/r$ Coulomb interactions \cite{Khveshchenko2001,Khveshchenko2004,Hands2008,Drut2009,Son2007}.
As a powerful method which enables us to treat strong $1/r$ Coulomb interactions properly, the U(1) lattice gauge theory has been applied to discuss the semimetal-insulator transition in graphene \cite{Drut2009,Drut2009a,Araki2010,Armour2011,Giedt2011,Buividovich2012}.
In this theory, the value of the chiral condensate is used as the order parameter for the transition.
It should be noted that the value obtained in an analytical calculation, the strong coupling expansion of the lattice gauge theory \cite{Araki2010}, and the value obtained in a numerical calculation \cite{Drut2009,Drut2009a} are in good agreement in the strong coupling region.

In this paper, we focus on the effects of strong $1/r$ long-range Coulomb interactions in multinode Dirac semimetals.
Due to the vanishing density of states near the Fermi level, the screening effect is considered to be weak in Dirac fermion systems.
Then it is expected that long-range interactions become important.
Based on the U(1) lattice gauge theory, we introduce two effective lattice models and take into account $1/r$ long-range Coulomb interactions between the bulk electrons.
Further we take into account the out-of-plane Fermi velocity anisotropy of the Dirac cones, since it is not small in experimentally observed Dirac semimetals \cite{Liu2014,Neupane2014}.
With the use of the strong coupling expansion of the lattice gauge theories and the mean-field approximation, we analyze the system from the strong coupling limit.
The value of the chiral condensate, which is equivalent to the dynamically generated mass of Dirac fermions, serves as the order parameter for the semimetal-insulator transition.

\section{Model}
Let us start from the effective continuum model for correlated $N$-node Dirac semimetals.
The model we consider is the (3+1)D four-component massless Dirac fermions of $N$ flavors interacting with the electromagnetic [U(1) gauge] field.
Compared to the usual quantum electrodynamics, our model is characterized by the Fermi velocity of Dirac fermions $v_{\rm F}$ which is much smaller than the speed of light $c$.
Due to this nature, the interactions via the vector potential (spatial components of the electromagnetic field) is suppressed by  the factor $v_{\rm F}/c\sim 10^{-3}$.
Then the Euclidean action of the system can be written as
\begin{align}
\begin{split}
S=&\int d^4x \sum_{f=1}^N\bar{\psi}_f(x) \left[\gamma_0(\partial_0+iA_0)+\xi_j\gamma_j\partial_j\right]\psi_f(x)\\
&+\frac{\beta}{2}\int d^4x (\partial_i A_0)^2,\label{continuum-action}
\end{split}
\end{align}
where $\psi_f(x)$ is a four-component spinor with $f$ denoting the flavor of Dirac fermions, $\gamma_\mu$ ($\mu=0,1,2,3$) are the $4\times4$ gamma matrices which satisfy the Clifford algebra $\{\gamma_\mu,\gamma_\nu\}=2\delta_{\mu\nu}$, and $A_0$ is the scalar potential.
Here we have introduced parameters for the Fermi velocity anisotropy $\xi_j$ with $\xi_1=\xi_2=1$ and $\xi_3=v_{\rm F\perp}/v_{\rm F \parallel}$.
Note that we have rescaled variables as $v_{\rm F\parallel}x_0\rightarrow x_0$, $A_0/v_{\rm F\parallel}\rightarrow A_0$ in Eq. (\ref{continuum-action}).
A parameter $\beta$, which represents the effective strength of the $1/r$ Coulomb interactions, is given by
\begin{align}
\begin{split}
\beta=\frac{v_{\rm F\parallel}\epsilon}{e^2}=\frac{v_{\rm F\parallel}\epsilon}{4\pi c \alpha},
\end{split}
\end{align}
where $e$ is the electric charge, $\epsilon$ is the dielectric constant of the system, and $\alpha(\simeq 1/137)$ is the fine-structure constant.
The smallness of the Fermi velocity makes the Coulomb interactions effectively strong. 
$\beta=0$ corresponds to the strong coupling limit.
In this study we consider the case of $\beta\ll 1$, i.e. the case of small dielectric constant. 

In the following, we introduce two specific effective lattice models for $N$-node Dirac semimetals with $N=4$ and $N=16$.
We take advantage of the so-called ``fermion doubling problem'' which occurs when considering Dirac fermions on lattices.
It is known that the fermion doublers can emerge in the cases where lattice fermions possess chiral symmetry, which has been proved by the Nielsen-Ninomiya theorem \cite{NN-theorem}.

{\it The $N=16$ Dirac Semimetal.---}
First we consider a (3+1)D $N=16$ Dirac semimetal interacting via $1/r$ Coulomb interactions on a lattice.
As the noninteracting action, we adopt the doubled lattice fermions (in the chiral limit) which reproduce the four-component massless Dirac fermions of 16 flavors in the continuum limit \cite{Wilson1974}.
The Euclidean action of the system is given by $S^{(N=16)}=S^{(N=16)}_F+S_{G}$. The fermionic part $S^{(N=16)}_F$ is written as
\begin{align}
\begin{split}
S^{(N=16)}_F=&\frac{1}{2}\sum_{n}\left[\bar{\psi}_n\gamma_0 U_{n,0}\psi_{n+\hat{0}} - \bar{\psi}_{n+\hat{0}}\gamma_0 U^\dag_{n,0}\psi_n\right]\\
&+\frac{1}{2}\sum_{n,j}\xi_j\left[\bar{\psi}_n\gamma_j\psi_{n+\hat{j}} - \bar{\psi}_{n+\hat{j}}\gamma_j\psi_n\right],
\end{split}\label{S_Wilson}
\end{align}
where $\psi_n$ is a four-component spinor.
This action is understood as the naively discretized action of the four-component Dirac fermions of single flavor.
The U(1) gauge part $S_G$ is written as
\begin{align}
\begin{split}
S_G=\beta\sum_n\sum_{\mu>\nu}\left[1-\frac{1}{2}\left(U_{n,\mu}U_{n+\hat{\mu},\nu}U^\dag_{n+\hat{\nu},\mu}U^\dag_{n,\nu}+{\rm H.c.}\right)\right].
\end{split}\label{S_G}
\end{align}
Here $\hat{\mu}$ ($\mu=0,1,2,3$) denotes the unit vector along the $\mu$ direction, and
$n=(n_0,n_1,n_2,n_3)$ is a lattice site on a four-dimensional isotropic lattice.
The U(1) gauge link variables $U_{n,\mu}$ are given by $U_{n,0}=e^{iA_0(n)}\equiv e^{i\theta_n}$ ($-\pi\le\theta_n\le\pi$) and $U_{n,j}=1$.

{\it The $N=4$ Dirac Semimetal.---}
Next we consider a (3+1)D $N=4$ Dirac semimetal interacting via $1/r$ Coulomb interactions on a lattice.
As the noninteracting action, we adopt the staggered fermions (in the chiral limit) which reproduce the four-component massless Dirac fermions of $4$ flavors in the continuum limit \cite{Kogut1975,Susskind1977}.
The Euclidean action of the system is given by $S^{(N=4)}=S^{(N=4)}_F+S_{G}$. The fermionic part $S^{(N=4)}_F$ is written as
\begin{align}
\begin{split}
S^{(N=4)}_F&=\frac{1}{2}\sum_{n}\eta_{n,0}\left[\bar{\chi}_nU_{n,0}\chi_{n+\hat{0}} - \bar{\chi}_{n+\hat{0}} U^\dag_{n,0}\chi_n\right]\\
&\quad+\frac{1}{2}\sum_{n,j}\xi_j\eta_{n,j}\left[\bar{\chi}_n\chi_{n+\hat{j}} - \bar{\chi}_{n+\hat{j}}\chi_n\right],
\end{split}\label{S_staggered}
\end{align}
where $\chi_n$ is a single-component spinor, $\eta_{n,0}=1$, $\eta_{n,1}=(-1)^{n_0}$, $\eta_{n,2}=(-1)^{n_0+n_1}$, and $\eta_{n,3}=(-1)^{n_0+n_1+n_2}$.
The gauge part $S_G$ is the same as Eq. (\ref{S_G}).
The action (\ref{S_staggered}) can be understood as an action obtained by doing the spin diagonalization (the Kawamoto-Smit transformation) \cite{Kawamoto1981} to $\psi_n$ in the action (\ref{S_Wilson}) as
\begin{align}
\begin{split}
\psi_n=T_n\xi_n, \ \ \ \ \ \bar{\psi}_n=\bar{\xi}_n T^\dagger_n
\end{split}
\end{align}
with $T_n=(\gamma_0)^{n_0}(\gamma_1)^{n_1}(\gamma_2)^{n_2}(\gamma_3)^{n_3}$ and $\xi_n\equiv[\chi_n^1,\chi_n^2,\chi_n^3,\chi_n^4]^T$, and then by retaining one of the four components in $\xi_n$.
However, to be precise, the action of staggered fermions after recovering the spinor structure does not coincide with that of Wilson fermions.
This is known as the taste breaking of staggered fermions. 
The (2+1)D staggered fermions have been used as an effective lattice model for graphene \cite{Drut2009,Drut2009a,Araki2010,Armour2011,Giedt2011,Buividovich2012}, since they reproduce the four-component massless Dirac fermions of 2 flavors in the continuum limit \cite{Burden1987}.

\section{Strong coupling expansion}
Let us derive the effective actions in the strong coupling limit ($\beta=0$). We can derive the effective action $S_{\rm eff}$ by integrating out the U(1) gauge link variable $U_{0,n}$ in the partition function $Z$ up to the arbitrary order in $\beta$ as follows:
\begin{align}
\begin{split}
Z^{(N=16)}=\int \mathcal{D}[\psi,\bar{\psi},U_0]e^{-S^{(N=16)}}=\int \mathcal{D}[\psi,\bar{\psi}]e^{-S^{(N=16)}_{\rm eff}}.\label{partition-function}
\end{split}
\end{align}
Here we have written down the case of the $N=16$ Dirac semimetal explicitly.
The same method can be applied to the case of $N=4$ by replacing $\psi$ to $\chi$.
In the strong coupling limit, $S_G$ vanishes and thus $U_{n,0}$ is contained only in $S^{(N=16)}_F$ and $S^{(N=4)}_F$.
Then the integral $\int \mathcal{D}U_0e^{-S^{(N=16)}_F}$ is performed as
\begin{align}
\begin{split}
&\prod_n\int_{-\pi}^\pi\frac{d\theta_n}{2\pi}\exp\left\{\frac{1}{2}\left[\bar{\psi}_n\gamma_0 U_{n,0}\psi_{n+\hat{0}} - \bar{\psi}_{n+\hat{0}}\gamma_0 U^\dag_{n,0}\psi_n\right]\right\}\\
&=\prod_n\left[1-\frac{1}{4}\bar{\psi}_n\gamma_0 \psi_{n+\hat{0}}\bar{\psi}_{n+\hat{0}}\gamma_0 \psi_n+\cdots\right]\\
&\approx e^{\frac{1}{4}\sum_n\mathrm{tr}\left[\gamma_0^T\bar{\psi}_n\psi_n\gamma_0^T\bar{\psi}_{n+\hat{0}}\psi_{n+\hat{0}}\right]},
\end{split}\label{U0-integration}
\end{align}
where we have used the fact that the Grassmann variables $\psi_\alpha$ and $\bar{\psi}_\alpha$ satisfy $\psi_\alpha^2=\bar{\psi}_\alpha^2=0$ with $\alpha$ denoting the component of the spinors.
In the second line, we have neglected the terms which consist of $8,12$ and $16$ different Grassmann variables.
As is mentioned in Sec. \ref{Results}, their contributions appear in higher orders of the order parameter, and do not affect the discussion on the semimetal-insulator transition in this model.
Further in the last line, we have rewritten the exponent as
\begin{align}
\begin{split}
\bar{\psi}_{n,\alpha}(\gamma_0)_{\alpha\beta} \psi_{n+\hat{0},\beta}\bar{\psi}_{n+\hat{0},\gamma}(\gamma_0)_{\gamma\delta} \psi_{n,\delta}\\
= -\mathrm{tr}\left[\gamma_0^T\bar{\psi}_n\psi_n\gamma_0^T\bar{\psi}_{n+\hat{0}}\psi_{n+\hat{0}}\right].
\end{split}
\end{align}
The subscripts $\alpha$ and $\beta$ denote the component of the spinors, and the superscript $T$ denotes the transpose of a matrix.
In the general cases of SU($N_c$) gauge field ($N_c\ge1$), we can perform the integration with respect to the gauge link variables $U$ in Eq. (\ref{partition-function}) by using the SU($N_c$) group integral formulas: $\int dU 1=1$, $\int dU U_{ab}=0$, $\int dU U_{ab}U_{cd}^\dagger=\delta_{ad}\delta_{bc}/N_c$, and so on.
Finally we obtain the effective action of the $N=16$ Dirac semimetal in the strong coupling limit given by
\begin{align}
\begin{split}
S^{(N=16)}_{{\rm eff}}&=\frac{1}{2}\sum_{n,j}\xi_j\left[\bar{\psi}_n\gamma_j\psi_{n+\hat{j}} - \bar{\psi}_{n+\hat{j}}\gamma_j\psi_n\right]\\
&\quad-\frac{1}{4}\sum_n\mathrm{tr}\left[\gamma_0^T\bar{\psi}_n\psi_n\gamma_0^T\bar{\psi}_{n+\hat{0}}\psi_{n+\hat{0}}\right].\label{Effective-action1}
\end{split}
\end{align}
From this equation, we see that the electron-electron interactions in the strong coupling limit is spatially on-site interaction but not in the (imaginary) time.

In the $N=4$ case, $\chi$ is a single-component Grassmann variable.
Therefore, due to the nature of Grassmann variables $\chi^2=\bar{\chi}^2=0$,  the approximation done in the second line of Eq. (\ref{U0-integration}) is not needed.
Finally we obtain the effective action of the $N=4$ Dirac semimetal in the strong coupling limit as
\begin{align}
\begin{split}
S^{(N=4)}_{{\rm eff}}&=\frac{1}{2}\sum_{n,j}\xi_j\eta_{n,j}\left[\bar{\chi}_n\chi_{n+\hat{j}} - \bar{\chi}_{n+\hat{j}}\chi_n\right]\\
&\quad-\frac{1}{4}\sum_n\bar{\chi}_n\chi_n\bar{\chi}_{n+\hat{0}}\chi_{n+\hat{0}}.\label{Effective-action2}
\end{split}
\end{align}
Note that this action is exact in the strong coupling limit, although we call it ``effective action''.

\section{Free Energies in the strong coupling limit \label{FreeEnergy}}
Let us derive the free energies in the strong coupling limit at zero temperature.
To this end, we apply the extended Hubbard-Stratonovich transformation to the interaction terms.
First let us consider the case of the $N=16$ Dirac semimetal.
In this case, introducing the two complex matrix auxiliary fields $Q$ and $Q'$, $e^{\kappa\mathrm{tr}AB}$ with $\kappa>0$ and $A,B$ being matrices is deformed as follows \cite{Sekine2013a}:
\begin{align}
\begin{split}
&e^{\kappa\mathrm{tr}AB}={\rm (const.)}\times\\
&\int \mathcal{D}[Q,Q'] \exp\left\{-\kappa\left[Q_{\alpha\beta}Q'_{\alpha\beta}-A_{\alpha\beta}Q_{\beta\alpha}-B^T_{\alpha\beta}Q'_{\beta\alpha}\right]\right\},\label{EHS}
\end{split}
\end{align}
This integral is approximated by the saddle point values $Q_{\alpha\beta}=\langle B^T\rangle_{\beta\alpha}$ and $Q'_{\alpha\beta}=\langle A\rangle_{\beta\alpha}$.
In the case of the $N=4$ Dirac semimetal, we can apply Eq. (\ref{EHS}) with the subscripts removed, since there is no spinor structure in the action.

{\it Free Energy of the $N=16$ Dirac Semimetal.---}
We set $(\kappa, A, B)=(1/4, \gamma_0^T\bar{\psi}_n\psi_n, \gamma_0^T\bar{\psi}_{n+\hat{0}}\psi_{n+\hat{0}})$ to decouple the interaction term (the second term) of Eq. (\ref{Effective-action1}) to fermion bilinear form.
In this process, we need to assume the form of the $4\times 4$ matrix $\langle \bar{\psi}_n\psi_n\rangle$ by the mean-field approximation.
Recall that the purpose of this study is to discuss the semimetal-insulator transition induced by strong long-range Coulomb interactions.
Here let us consider the possible gapped phases in our model.
In the action (\ref{S_Wilson}) with $U_{n,0}=1$, only the identity matrix $\bm{1}$ and the matrix $\gamma_5$ can open energy gaps.
This is because, in the presence of these matrices, the single-particle Hamiltonian of the system is given by
\begin{align}
\begin{split}
\mathcal{H}(\bm{k})=\xi_j\alpha_j\sin k_j+m_4\alpha_4+m_5\alpha_5
\end{split}\label{MF-Hamiltonian}
\end{align}
with $\alpha_j=\gamma_0\gamma_j$, $\alpha_4=\gamma_0$ and $\alpha_5=i\gamma_0\gamma_5$, which leads to the gapped energy spectrum $E(\bm{k})=\pm\sqrt{\sum_j(\xi_j\sin k_j)^2+m_4^2+m_5^2}$.
Note that the action (\ref{S_Wilson}) possesses chiral symmetry, namely, the action is invariant under the chiral transformation $\psi_n\rightarrow e^{i\alpha\gamma_5}\psi_n$.
In such cases, as in the case of graphene \cite{Hands2008,Drut2009,Drut2009a,Araki2010,Armour2011,Giedt2011,Buividovich2012}, the identity matrix (i.e. the mass term) serves as the order parameter for the semimetal-insulator transition.
Therefore we can set $\langle\bar{\psi}_n\psi_n\rangle=-\sigma\bm{1}$.
If the value of $\sigma$ is nonzero in the strong coupling limit, then the value of $\sigma$ corresponds to $\sigma=\sqrt{m_4^2+m_5^2}$ in the energy spectrum $E(\bm{k})$.
Namely, we obtain the gapped spectrum.
We can regard the value of $\sigma$ as the dynamically generated mass of Dirac fermions.
In the lattice QCD, $\sigma$ is known as the ``chiral condensate''.

Then the terms $Q_{\alpha\beta}Q'_{\alpha\beta}$, $A_{\alpha\beta}Q_{\beta\alpha}$ and $B^T_{\alpha\beta}Q'_{\beta\alpha}$ in the integrand of Eq. (\ref{EHS}) are calculated explicitly as
\begin{align}
\begin{split}
Q_{\alpha\beta}Q'_{\alpha\beta}&=\langle B^T\rangle_{\beta\alpha}\langle A\rangle_{\beta\alpha}={\rm tr}\left[\langle B\rangle\langle A\rangle\right]=\sigma^2{\rm tr}\left[(\gamma_0^T)^2\right]\\
&=4\sigma^2,
\end{split}
\end{align}
\begin{align}
\begin{split}
A_{\alpha\beta}Q_{\beta\alpha}&=A_{\alpha\beta}\langle B^T\rangle_{\alpha\beta}={\rm tr}\left[A\langle B\rangle\right]
=-\sigma{\rm tr}\left[\gamma_0^T\bar{\psi}_n\psi_n\gamma_0^T\right]\\
&=-\sigma\bar{\psi}_n\psi_n,
\end{split}
\end{align}
\begin{align}
\begin{split}
B^T_{\alpha\beta}Q'_{\beta\alpha}&=B^T_{\alpha\beta}\langle A\rangle_{\alpha\beta}={\rm tr}\left[B\langle A\rangle\right]
=-\sigma{\rm tr}\left[\gamma_0^T\bar{\psi}_n\psi_n\gamma_0^T\right]\\
&=-\sigma\bar{\psi}_n\psi_n,
\end{split}
\end{align}
where we have used $\langle\bar{\psi}_{n+\hat{0}}\psi_{n+\hat{0}}\rangle=\langle\bar{\psi}_{n-\hat{0}}\psi_{n-\hat{0}}\rangle=\langle\bar{\psi}_n\psi_n\rangle=-\sigma\bm{1}$ and $(\gamma_0^T)^2=(\gamma_0^2)^T=\bm{1}$.
Combining these three equations, we obtain the interaction term decoupled to fermion bilinear:
\begin{align}
\begin{split}
\frac{1}{4}\sum_n\mathrm{tr}\left[\gamma_0^T\bar{\psi}_n\psi_n\gamma_0^T\bar{\psi}_{n+\hat{0}}\psi_{n+\hat{0}}\right]\sim -\frac{1}{4}\sum_{n}\left[4\sigma^2+2\sigma\bar{\psi}_n\psi_n\right].
\end{split}\label{int-term1}
\end{align}
We are now in a position to derive the free energy at zero temperature in the strong coupling limit.
Combining Eqs. (\ref{Effective-action1}) and (\ref{int-term1}), the effective action expressed by the auxiliary field $\sigma$ is given by
\begin{align}
\begin{split}
S^{(N=16)}_{\mathrm{eff}}(\sigma)=\frac{V}{T}\sigma^2+\sum_{k=-\pi}^\pi \bar{\psi}_k\mathcal{M}(\bm{k};\sigma)\psi_k \label{S_eff_SCL}
\end{split}
\end{align}
with $\mathcal{M}(\bm{k};\sigma)=\sum_j\xi_ji\gamma_j\sin k_j+\sigma/2$.
As mentioned above, the value $\sigma/2$ can be regarded as the dynamically generated mass.
Here $V$ and $T$ are the volume and temperature of the system, respectively, and we have done the Fourier transform from $n=(n_0,\bm{n})$ to $k=(k_0,\bm{k})$.
From this action, we derive the free energy at zero temperature per unit spacetime volume $\mathcal{F}^{(N=16)}(\sigma)$, according to the usual formula $\mathcal{F}^{(N=16)}=-\frac{T}{V}\ln Z^{(N=16)}$.
The partition function $Z^{(N=16)}$ is calculated by the Grassmann integral formula $Z^{(N=16)}=\int D[\psi,\bar{\psi}]e^{-\bar{\psi}\mathcal{M}\psi}=\mathrm{det}\mathcal{M}$.
The determinant of $\mathcal{M}$ is calculated by the formula $\mathrm{det}\mathcal{M}=\sqrt{\mathrm{det}(\mathcal{M}\mathcal{M}^\dagger)}$.
Finally, after a straightforward calculation, we arrive at the free energy in the strong coupling limit:
\begin{align}
\begin{split}
\mathcal{F}^{(N=16)}(\sigma)=\sigma^2-2\int_{-\pi}^{\pi}\frac{d^3k}{(2\pi)^3}\ln\left[\sum_j\xi_j^2\sin^2 k_j+\frac{1}{4}\sigma^2\right].
\end{split}\label{F1-SCL}
\end{align}
The ground state is determined by the stationary condition $d \mathcal{F}^{(N=16)}(\sigma)/d \sigma=0$.

{\it Free Energy of the $N=4$ Dirac Semimetal.---}
We set $(\kappa, A, B)=(1/4, \bar{\chi}_n\chi_n, \bar{\chi}_{n+\hat{0}}\chi_{n+\hat{0}})$ to decouple the interaction term (the second term) of Eq. (\ref{Effective-action2}) to fermion bilinear form.
Like in the $N=16$ case above, we need to assume the value of $\langle \bar{\chi}_n\chi_n\rangle$ by the mean-field approximation.
Note that the lattice action (\ref{S_staggered}) also possesses chiral symmetry, namely, the action is invariant under the chiral transformation defined by $\chi_n\rightarrow e^{i\alpha\epsilon(n)}\chi_n$, $\bar{\chi}_n\rightarrow e^{i\alpha\epsilon(n)}\bar{\chi}_n$ with $\epsilon (n)=(-1)^{n_0+n_1+n_2+n_3}$.
Hence we can set $\langle \bar{\chi}_n\chi_n\rangle=-\sigma$.
Then with this approximation, the interaction term is decoupled to fermion bilinear as
\begin{align}
\begin{split}
\frac{1}{4}\sum_n\bar{\chi}_n\chi_n\bar{\chi}_{n+\hat{0}}\chi_{n+\hat{0}}\sim -\frac{1}{4}\sum_{n}\left[\sigma^2+2\sigma\bar{\chi}_n\chi_n\right].
\end{split}\label{int-term2}
\end{align}
It is convenient to perform the Fourier transform only to the spatial directions, due to the factor $\eta_{n,j}$ in the noninteracting part of the effective action (\ref{Effective-action2}).
By introducing the eight-component spinor $\Psi_{n_0,\bm{k}}$ as
\begin{align}
\Psi_{n_0,\bm{k}}=
\begin{bmatrix}
\chi_{n_0}(k_1,k_2,k_3)\\
\chi_{n_0}(k_1-\pi,k_2,k_3)\\
\chi_{n_0}(k_1,k_2-\pi,k_3)\\
\chi_{n_0}(k_1,k_2,k_3-\pi)\\
\chi_{n_0}(k_1,k_2-\pi,k_3-\pi)\\
\chi_{n_0}(k_1-\pi,k_2,k_3-\pi)\\
\chi_{n_0}(k_1-\pi,k_2-\pi,k_3)\\
\chi_{n_0}(k_1-\pi,k_2-\pi,k_3-\pi)
\end{bmatrix},
\end{align}
and substituting Eq. (\ref{int-term2}) to Eq. (\ref{Effective-action2}), the effective action is rewritten as
\begin{align}
\begin{split}
S^{(N=4)}_{\mathrm{eff}}(\sigma)=\frac{1}{4}\frac{V}{T}\sigma^2+\sum_{n_0}\sum_{\bm{k}=0}^\pi \bar{\Psi}_{n_0,\bm{k}}^T\mathcal{V}(n_0,\bm{k};\sigma)\Psi_{n_0,\bm{k}}.
\end{split}
\end{align}
Note that the sum over the wave vector $k_j$ is from $0$ to $\pi$.
The procedure to derive the free energy at zero temperature per unit spacetime volume $\mathcal{F}^{(N=4)}(\sigma)$ is the same as the case of $N=16$.
The calculation of $\mathrm{det}\mathcal{V}$ is a little complicated but can be done analytically to be $\mathrm{det}\mathcal{V}=[\sum_j\xi_j^2\sin^2 k_j+\frac{1}{4}\sigma^2]^4$.
Finally we arrive at the free energy in the strong coupling limit:
\begin{align}
\begin{split}
\mathcal{F}^{(N=4)}(\sigma)=\frac{1}{4}\sigma^2-\frac{1}{2}\int_{-\pi}^{\pi}\frac{d^3k}{(2\pi)^3}\ln\left[\sum_j\xi_j^2\sin^2 k_j+\frac{1}{4}\sigma^2\right],
\end{split}\label{F2-SCL}
\end{align}
where we have used the fact that the integrand is an even function.
The ground state is determined by the stationary condition $d \mathcal{F}^{(N=4)}(\sigma)/d \sigma=0$.

\section{Numerical Results \label{Results}}
\begin{figure}[!t]
\centering
\includegraphics[width=1.0\columnwidth,clip]{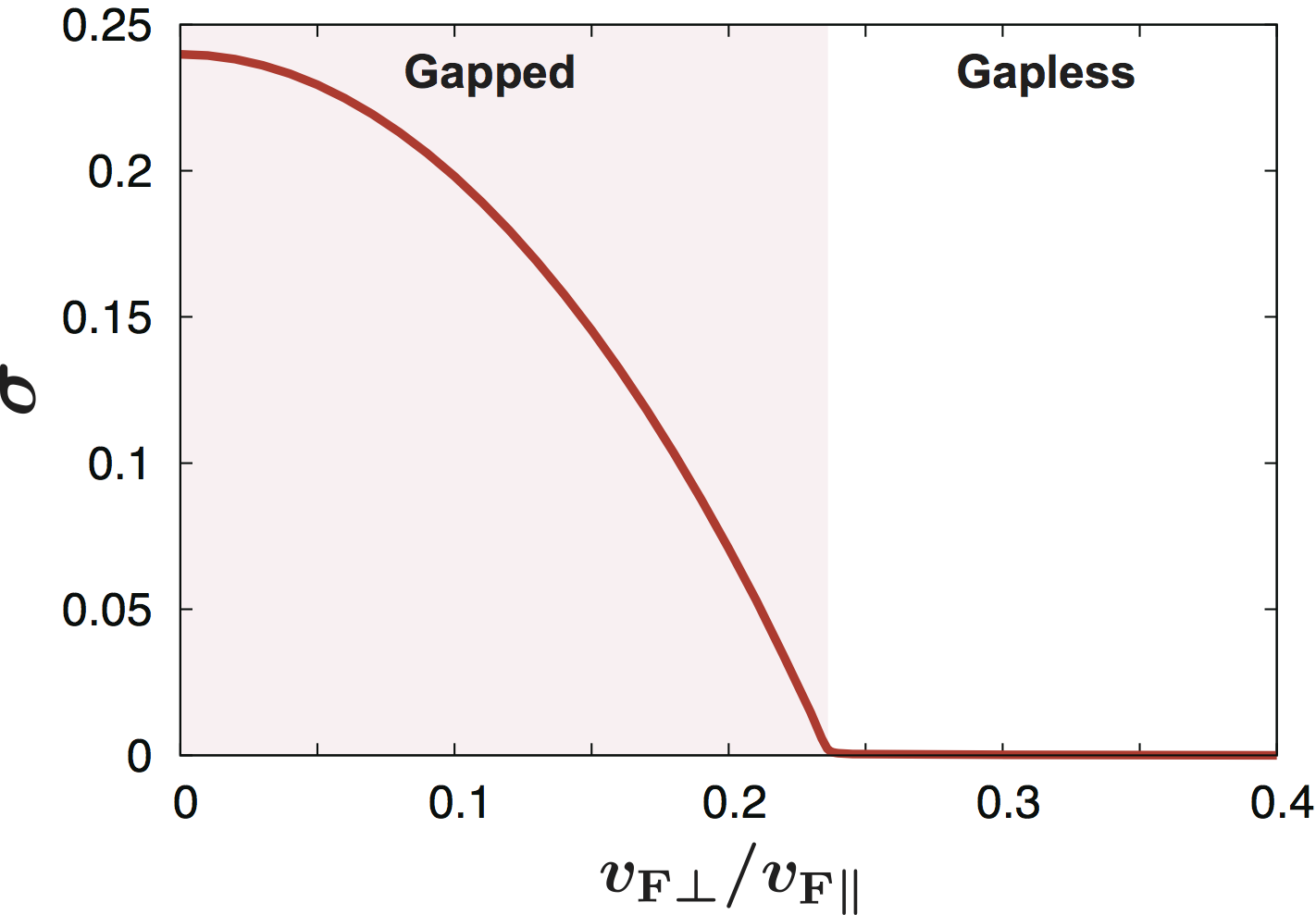}
\caption{(Color online) Fermi velocity anisotropy $v_{\rm F\perp}/v_{\rm F \parallel}$ dependence of the chiral condensate $\sigma$ for the $N=4$ case in the strong coupling limit ($\beta=0$). When $v_{\rm F\perp}/v_{\rm F \parallel}=0$, the system corresponds to monolayer graphene in the strong coupling limit.}\label{Fig1}
\end{figure}

First we consider the result for the $N=4$ case.
The Fermi velocity anisotropy $v_{\rm F\perp}/v_{\rm F \parallel}(=\xi_3)$ dependence of the chiral condensate $\sigma$ in the strong coupling limit is shown in Fig. \ref{Fig1}.
It was found that the the value of $\sigma$ becomes zero when the ratio $v_{\rm F\perp}/v_{\rm F \parallel}$ is larger than about $0.24$, whereas the value of $\sigma$ is nonzero when $v_{\rm F\perp}/v_{\rm F \parallel}$ is smaller than about $0.24$.
As mentioned in Sec. \ref{FreeEnergy}, the value of $\sigma$, the chiral condensate, is regarded as the dynamically generated mass of Dirac fermions, and can be used as the order parameter for the semimetal-insulator transition.
Hence the result indicates that whether the system is insulating or semimetallic (i.e. gapped or gapless) in the strong coupling limit depends on the value of the Fermi velocity anisotropy.
Namely, the $N=4$ Dirac semimetals survive in the strong coupling limit when the anisotropy is weak, whereas they change to Mott insulators when the anisotropy is strong.
We see from Fig. \ref{Fig1} that the transition is of the second order.

The result, that the system becomes gapped in the strong coupling limit when the Fermi velocity anisotropy is strong (i.e. the ratio $v_{\rm F\perp}/v_{\rm F \parallel}$ is small), could be understood by the fact that in general quantum effects become stronger in lower dimensions.
In the case of monolayer graphene, theoretical studies have shown that the graphene suspended in vacuum (or the graphene on a substrate with sufficiently small dielectric constant) becomes gapped due to strong $1/r$ Coulomb interactions \cite{Khveshchenko2001,Khveshchenko2004,Khveshchenko2007,Hands2008,Drut2009,Drut2009a,Araki2010,Armour2011,Giedt2011,Buividovich2012}.
In the the $N=4$ case, the interaction term in the effective action [Eq. (\ref{Effective-action2})] describes spatially on-site interactions.
This means that our model with $v_{\rm F\perp}=0$ in the strong coupling limit is equivalent to a model for a stacked 2D system.
To be more precise, our model with $v_{\rm F\perp}=0$ in the strong coupling limit corresponds to an effective lattice model for monolayer graphene in the strong coupling limit, since the (2+1)D staggered fermions reproduce the four-component Dirac fermions of 2 flavors in the continuum limit \cite{Burden1987}.
Actually, the value of $\sigma$ when $v_{\rm F\perp}=0$ in our model, $\sigma\simeq 0.24$, is equal to the value obtained by a lattice strong coupling expansion study of monolayer graphene \cite{Araki2010}.

Here we would like to mention the correctness of our value of $\sigma$ in the strong coupling limit.
As for monolayer graphene which is described by (2+1)D staggered fermions, the value of $\sigma$ obtained in a lattice strong coupling expansion study with the mean-field approximation \cite{Araki2010} is in qualitative agreement (within about 10$\%$ of difference) with the values obtained in Monte Carlo studies \cite{Drut2009,Drut2009a}.
Hence it is expected that our value of $\sigma$ for the $N=4$ case is quantitatively correct, because the mean-field approximation gives more proper results in higher dimensions in general.

Finally we consider the result for the $N=16$ case.
We see easily that $\mathcal{F}^{(N=16)}(\sigma)=4\mathcal{F}^{(N=4)}(\sigma)$.
Namely, within our calculation the values of $\sigma$ for both the $N=4$ and the $N=16$ cases, which serve as the order parameter for the semimetal-insulator transition, are equivalent in the strong coupling limit.
Here note that we have neglected the interaction terms which consist of $8,12$ and $16$ fermion fields when deriving the effective action of the $N=16$ Dirac semimetal [Eq. (\ref{Effective-action1})].
If the 8 fermion field term, $S_8\sim \bar{\psi}\psi\bar{\psi}\psi\bar{\psi}\psi\bar{\psi}\psi$, is taken into account, then we obtain $S_8\sim \sigma^4+\sigma^3\bar{\psi}\psi$ by a rough mean-field approximation.
The 12 and 16 fermion field terms can be approximated in the same way.
When $\sigma=0$ at the stationary point of the free energy $\mathcal{F}^{(N=16)}(\sigma)$, i.e. when the Fermi velocity anisotropy is weak, the effects of these higher order terms in the free energy can be neglected near $\sigma=0$.
In other words, the result that the $N=16$ Dirac semimetal survives in the strong coupling limit will not be changed even though such terms are taken into account.

However, when $\sigma\neq 0$ at the stationary point of $\mathcal{F}^{(N=16)}(\sigma)$, i.e. when the anisotropy is strong, such terms will modify the value of $\sigma$ at the stationary point.
Here note that the interaction term in the effective action [Eq. (\ref{Effective-action1})] describes spatially on-site interactions.
Namely, our model with $v_{\rm F\perp}=0$ in the strong coupling limit corresponds to a model of stacked (2+1)D four-component Dirac fermions of 8 flavors in the strong coupling limit.
It has been reported in the (2+1)D cases that the value of $\sigma$ becomes smaller as the number of Dirac fermion flavor $N^{\rm 2D}$ becomes larger and the semimetal phase with large $N^{\rm 2D}$ survives in the strong coupling limit \cite{Hands2008,Drut2009,Son2007}.
Therefore, when we take into account those higher order terms in the free energy, it is expected that the value of $\sigma$ is suppressed in the case of small $v_{\rm F\perp}/v_{\rm F \parallel}$.
To verify this prediction, further study is needed.

\section{Discussions}
Firstly, we note the relations between our models and the experimentally observed Dirac semimetals.
In the observed Dirac semimetals, there exist large out-of-plane Fermi velocity anisotropy such that $v_{\rm F\perp}/v_{\rm F \parallel}\approx 0.25$ in Na$_3$Bi \cite{Liu2014} and $v_{\rm F\perp}/v_{\rm F \parallel}\sim 0.1$ in Cd$_3$As$_2$ \cite{Neupane2014}.
Therefore we expect that our result gives some perception to realistic materials.
On the other hand, as for the number of the Dirac nodes $N$, $N$ is two in both Na$_3$Bi and Cd$_3$As$_2$.
From the results in the (2+1)D case, i.e. multilayer graphene \cite{Drut2009}, it is expected that the dynamically generated mass gap $\sigma$ in the strong coupling limit becomes larger with decreasing $N$.
However, it is difficult to show such a behavior explicitly in our study.
Hence the stability of the $N=2$ case is a remaining problem.

\begin{figure}[!t]
\centering
\includegraphics[width=1.0\columnwidth,clip]{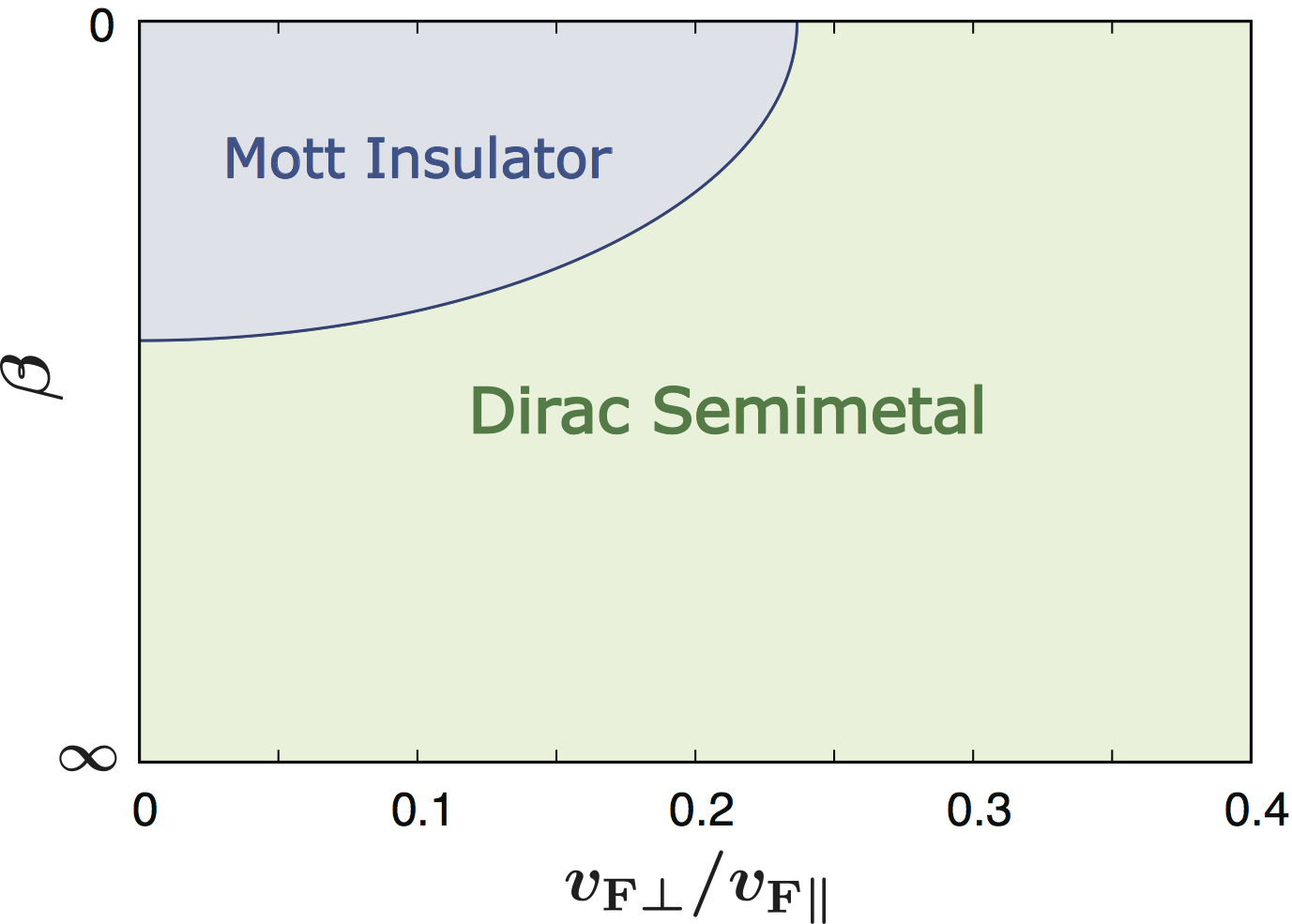}
\caption{(Color online) A possible global phase diagram of a correlated $N$-node Dirac semimetal with $N=4$.
The $\beta=0$ ($\beta=\infty$) line represents the strong coupling limit (noninteracting limit).
The Mott insulator phase is defined as the phase with nonzero value of $\sigma$.}\label{Fig2}
\end{figure}

Secondly, let us discuss a possible global phase diagram of correlated $N$-node Dirac semimetals with $N=4$ and $N=16$.
We see from Fig. \ref{Fig1} that as the ratio $v_{\rm F\perp}/v_{\rm F \parallel}$ is increased from zero, the value of $\sigma$ becomes smaller and eventually reaches zero.
The chiral condensate $\sigma$ can be used as the order parameter for the semimetal-insulator transition.
Namely, the system is gapless in the strong coupling limit ($\beta=0$) when the ratio $v_{\rm F\perp}/v_{\rm F \parallel}$ is large, whereas the system is gapped when the ratio $v_{\rm F\perp}/v_{\rm F \parallel}$ is small.
We call the gapped phase with nonzero $\sigma$ the Mott insulator.
On the other hand, the system is obviously a Dirac semimetal in the noninteracting limit ($\beta=\infty$).
Therefore, there must exist the critical strength of the $1/r$ Coulomb interactions $\beta_c$, below which the system becomes semimetallic, i.e. the value of $\sigma$ becomes zero.
This critical value $\beta_c$ will become smaller as the value of $\sigma$ in the strong coupling limit becomes smaller.
A schematic global phase diagram for the $N=4$ case based on this analysis is shown in Fig. \ref{Fig2}.
In the case of $N=16$, as mentioned in the previous section, it is expected that the value of $\sigma$ in the strong coupling limit is suppressed when the ratio $v_{\rm F\perp}/v_{\rm F \parallel}$ is small.
Namely, it is expected that the region of the Mott insulator phase shrinks in the global phase diagram.

In this study, we have focused only on the energy gap generation by strong $1/r$ Coulomb interactions, and thus the detailed information of the spinors in the low-energy effective model [Eq. (\ref{continuum-action})] is not required.
However, if we construct a low-energy effective model of some realistic material, then the spinors should be associated with the lattice structure and the spins of electrons, as in the case of graphene.
Hence it is expected that some order such as a magnetic or charge order is realized in the Mott insulator phase in Fig. \ref{Fig2}, although it is difficult in this study to identify what the order is.
This can be understood from the fact that the two possible orders in the Hamiltonian (\ref{MF-Hamiltonian}), the $\alpha_4$ order where no symmetry is broken and the $\alpha_5$ order where both time-reversal and inversion symmetries are broken, are energetically degenerate.
The result of this paper is not changed even though the $\alpha_5$ order is considered as the gapped order instead of the $\alpha_4$ order.

Thirdly, let us discuss the implications of our results to the stability of Weyl semimetals.
Weyl semimetals have gapless 3D linear dispersions in the bulk, and the effective Hamiltonians around the Weyl nodes are described by not the Dirac Hamiltonian but the Weyl Hamiltonian $\mathcal{H}_{\rm Weyl}(\bm{k})=\sum_{i=1}^3\bm{v}_i\cdot\bm{k}\sigma_i$ where $\sigma_i$ are the Pauli matrices.
At least either time-reversal or inversion symmetry breaking is required to realize a Weyl semimetal phase \cite{Volovik2003,Volovik2007,Murakami2007,Wan2011,Burkov2011a,Burkov2011,Kurebayashi2014,Halasz2012,Zyuzin2012a}.
Each Weyl node possesses chirality defined by $c={\rm sgn}[\bm{v}_1\cdot(\bm{v}_2\times\bm{v}_3)]=\pm1$. The number of the Weyl nodes with chirality $+1$ and that of chirality $-1$ must be equal in time-reversal symmetry broken Weyl semimetals.
For example, the Weyl semimetal phase predicted by a first-principles calculation in pyrochlore iridates possesses 24 Weyl nodes \cite{Wan2011}.
Here let us consider a simplified low-energy effective model for a Weyl semimetal with $2N$ nodes.
The Hamiltonian of such a system can be written as
\begin{align}
\begin{split}
H_{\rm Weyl}^{\rm eff}&=\sum_{\bm k}\sum_{f=1}^Nv_{{\rm F}\parallel}\left\{\psi^\dagger_{f+}(\bm{k})\left[\xi_ik_i\sigma_i\right]\psi_{f+}(\bm{k})\right.\\
&\quad\left.+\psi^\dagger_{f-}(\bm{k})\left[-\xi_ik_i\sigma_i\right]\psi_{f-}(\bm{k})\right\}\\
&=\sum_{\bm k}\sum_{f=1}^N\psi^\dagger_{f}(\bm{k})v_{{\rm F}\parallel}
\begin{bmatrix}
\xi_ik_i\sigma_i & 0\\
0 & -\xi_ik_i\sigma_i
\end{bmatrix}
\psi_{f}(\bm{k}),
\end{split}
\end{align}
where $\psi_f=[\psi_{f+},\psi_{f-}]^T$ with $\psi_{f\pm}$ being a two-component spinor, the subscript $\pm$ denotes the chirality of each Weyl node, and we have introduced the Fermi velocity anisotropy $\xi_i$.
By introducing the $4\times 4$ gamma matrices $\gamma_\mu$ in the chiral representation, we obtain the low-energy effective action (\ref{continuum-action}).
Namely, this indicates that in a rough approximation, the low-energy effective model of a $2N$-node Weyl semimetal is equivalent to that of a $N$-node Dirac semimetal. 
Weyl semimetals have a topological property such that the energy gap opens only if the Weyl nodes with opposite chirality meet each other, since a single Weyl fermion cannot be massive by itself.
Due to this property, Weyl semimetals are expected to be more stable against strong $1/r$ Coulomb interactions than Dirac semimetals.
However, it is not easy to treat strong $1/r$ Coulomb interactions properly in multinode Weyl semimetals.
In this study, it was found that the $N$-node Dirac semimetals with $N=4$ and $N=16$ survive in the strong coupling limit.
Hence, it could be said that the $N_{\rm W}$-node Weyl semimetals with $N_{\rm W}=8$ and $N_{\rm W}=32$ also survive in the strong coupling limit when the Fermi velocity anisotropy is weak.
As for the cases of $N_{\rm W}<8$, a recent study has reported that a Weyl semimetal with $N_{\rm W}=2$ survives in this limit \cite{Sekine2013b}.

Finally, let us discuss the implications of our results to the stability of 3D topological insulators.
It is known that 3D topological insulators can be regarded as 3D Dirac fermion systems.
In the noninteracting cases, the bulk energy gap closes when the phase transition from the topological insulator phase to the normal band insulator phase occurs.
In other words, there exist Dirac point(s) in the bulk when the system is on the phase boundary between the topological insulator phase and the normal band insulator phase.
For example, the Fu-Kane-Mele model has three Dirac points \cite{Fu2007,Fu2007a}, and the effective model for Bi$_2$Se$_3$ has one Dirac point \cite{Zhang2009} on their phase boundaries.
What about in the interacting cases?
The phase transitions from the topological insulator phase to the other phases can occur without the gap closing, when accompanying the breaking of symmetry of the system such as time-reversal symmetry or inversion symmetry.
However, the gap closing is required when no symmetry is broken, as in the noninteracting cases.
From this viewpoint, our result, that the Dirac semimetals survives in the strong coupling limit, suggests that 3D topological insulator phases can be stable against strong $1/r$ Coulomb interactions.
Actually, a recent study has reported that a 3D topological insulator phase of Bi$_2$Se$_3$-type survives in the strong coupling limit when the spin-orbit interaction of the system is strong \cite{Sekine2013a}.

\section{Summary}
In summary, based on the U(1) lattice gauge theory, we have studied the stability of $N$-node Dirac semimetals in three spatial dimensions with $N=4$ and $N=16$ against strong $1/r$ long-range Coulomb interactions.
It was shown that the Dirac semimetals survive in the strong coupling limit when the Fermi velocity anisotropy is weak, whereas they change to Mott insulators when the anisotropy is strong.
This means that the three-dimensionality of the Dirac cones plays an important role in the stability.
The value of the dynamically generated mass gap at least for the $N=4$ case is expected to be quantitatively correct.
A possible global phase diagram of correlated Dirac semimetals was presented.
Dirac semimetals can lead to various topological phases by the change of parameters or symmetry breakings.
Our result, that Dirac semimetals are stable against strong $1/r$ long-range Coulomb interactions, implies the stability of other topological phases.
Namely, it is suggested that Weyl semimetals, which correspond to Dirac semimetals in a rough approximation, can survive in the strong coupling limit.
The existence of 3D topological insulator phases in the strong coupling limit is also suggested.

\begin{acknowledgments}
A.S. is supported by the JSPS Research Fellowship for Young Scientists.
This work was supported by Grant-in-Aid for Scientific Research (No. 25103703, No. 26107505 and No. 26400308) from the Ministry of Education, Culture, Sports, Science and Technology (MEXT), Japan.
\end{acknowledgments}


\end{document}